\documentclass{ws-mpla}

\begin{document}

\markboth{Zhou L. L., B.-Q. Ma} {Lorentz Invariance Violation Matrix
from a General Principle}

\catchline{}{}{}{}{}

\title{Lorentz Invariance Violation Matrix from a General Principle\footnote{Published in Mod. Phys. Lett. A 25
(2010) 2489 - 2499}}

\author{ZHOU LINGLI}

\address{School of Physics
and State Key Laboratory of Nuclear Physics and Technology, \\
Peking University, Beijing 100871, China\\
zhoull@pku.edu.cn}

\author{BO-QIANG MA}

\address{School of Physics
and State Key Laboratory of Nuclear Physics and Technology, \\
Peking University, Beijing 100871, China\\
mabq@phy.pku.edu.cn}

\maketitle

\pub{Received (Day Month Year)}{Revised (Day Month Year)}

\begin{abstract}
We show that a general principle of physical independence
or physical invariance of mathematical background manifold leads to
a replacement of the common derivative operators by the covariant co-derivative ones.
This replacement naturally induces a background matrix, by means of which we obtain
an effective Lagrangian for the minimal standard model with
supplement terms characterizing Lorentz invariance violation
or anisotropy of space-time. We construct a simple model of the background matrix
and find that the strength of Lorentz violation of proton in the photopion production
is of the order $10^{-23}$.

\keywords{Lorentz invariance, Lorentz invariance violation matrix}
\end{abstract}

\ccode{PACS Nos.: 11.30.Cp, 03.70.+k, 12.60.-i, 01.70.+w}

\section{Introduction}

Lorentz symmetry is one of the most significant and fundamental
principles in physics, and it contains two aspects: Lorentz
covariance and Lorentz invariance. Nowadays, there have been
increasing interests in Lorentz invariance Violation (LV)
both theoretically and experimentally (see, e.g., Ref. \refcite{Wolfgang}). In this paper we find out a general
principle, which provides a consistent framework to describe
the LV effects. It requires the following replacements of the
ordinary partial derivative $\partial_{\alpha}$ and covariant derivative $D_{\alpha}$ by
the co-derivative ones
\begin{equation}\label{eqn:substitution}
\partial^{\alpha} \rightarrow M^{\alpha\beta}\partial_{\beta},\quad D^{\alpha}\rightarrow M^{\alpha\beta}D_{\beta},
\end{equation}
where $M^{\alpha\beta}$ is a local matrix. In the following, we introduce this
general principle at first, and then explore its physical implications and consequences.

\section{Principle of Physical Invariance}\label{sec_PIPC}

Principle: Under any one-to-one transformation $X\rightarrow X'=f(X)$ on
mathematical background manifold, the
transformation $\varphi(\cdot)\rightarrow \varphi'(\cdot)$ of an arbitrary physical field $\varphi(X)$ should satisfy
\begin{equation}\label{eqn:PFIMBM}
\varphi'(X')=\varphi(X).
\end{equation}

Most generally, this principle can be handled in geometric
algebra $ \mathcal{G}$ (or Clifford algebra) and geometric calculus
(see, e.g., Refs. \refcite{GAtoGC,Doran03}). The general element
in geometric algebra is called a multivector, and addition and various
products of two multivectors are still a multivector, i.e. geometric
algebra is closed. Different variables in physics, such as scalar,
vector, tensor, spinor, twistor, matrix, etc., can be described by the corresponding types of multivectors in a unified form
in geometric algebra (see, e.g., Refs. \refcite{GAtoGC,Doran03,Arcaute06,Matej06} for details).
$\varphi(X) \in \mathcal{G}$ is a multivector-valued function of a multivector
variable $X \in \mathcal{G}$, and it can be decomposed in an arbitrary local coordinate system with the basis
vectors $\{e_1,e_2,\ldots,e_n\}$ as
\begin{eqnarray}\label{MVdecompose}
\varphi(X)=\varphi^0+\sum_{\alpha=1}^{n}\varphi^{\alpha}e_{\alpha}+\cdots+\sum_{\alpha_1<\cdots<\alpha_{n}}\varphi^{\alpha_{1}\cdots \alpha_{n}}e_{\alpha_1}\wedge\cdots\wedge e_{\alpha_{n}}\equiv\sum_{J}\varphi^{J}e_{J},\nonumber
\end{eqnarray}
where ``$\wedge$'' is called wedge product, $e_J \equiv  e_{\alpha_{1}}\wedge\cdots\wedge e_{\alpha_{r}}$, with $J= \{\alpha_{1},\cdots,\alpha_{r}\}$ for $r=1,\cdots, n $, and $e_0\equiv 1$ with $J=0$.
Here $\varphi^{J}$ is the components of $\varphi(X)$ and $J$
denotes tensor index. We should not be frustrated by the
operation algorithms in geometric algebra, but could consider a
general multivector $\varphi$ just as an ordinary vector and that two multivectors
$\varphi$ and $\psi$ are not commutable, namely $\varphi\psi\neq\psi\varphi$, somehow like
matrices. Remembering these two points will suffice the following discussions.
What should be kept in mind is that in this new framework, the commonly used
tensor notation $\varphi^{\alpha_{1}\cdots\alpha_{n}}$ only represents the component of
$\varphi$ with respect to a specific coordinate basis. Conventionally one
simply means a multivector $\varphi$ by its component $\varphi^{\alpha_{1}\cdots \alpha_{n}}$
and regards the transformation rules for these components
from one basis to another as the transformation of a multivector. However, a multivector should be coordinate-independent.
For instance, $\varphi=\varphi^{\alpha}e_{\alpha}$ is a multivector and does not change under any coordinate transformation but
its component $\varphi^{\alpha}$ does.

This general principle actually makes the field $\varphi(X)$ represent a
physical distribution, rather than a common mathematical
function. Although the uniqueness of reality can be mathematically described in many ways like
$\varphi(X)$, $\varphi'(X')$, and $\varphi''(X'')$, $\cdots$, the physics behind remains unchanged, saying independence or
invariance. So (\ref{eqn:PFIMBM}) claims Physical Independence
or Physical Invariance (PI) on mathematical background manifold.

For the field $\varphi(X)$ satisfying (\ref{eqn:PFIMBM}), its derivative field might be naively defined as
\begin{equation}\label{def1}
\pi(X)=\partial_{X}\varphi(X).\nonumber
\end{equation}
But does this definition still fulfil (\ref{eqn:PFIMBM})? If $\pi(X)$ is a physical field, there must be the
condition $\pi'(X')=\pi(X)$ according to PI. However,
\begin{eqnarray}\label{eqn:momentum}
\pi(X)&=&\partial_{X}\varphi(X)=\partial_{X}\varphi'(X') \nonumber \\
      &=&\partial_{X}f(X)\ast\partial_{X'}\varphi'(X')\nonumber\\
      &=&F(\partial_{X'})\varphi'(X')\nonumber\\
      &\neq&\pi'(X'),\nonumber
\end{eqnarray}
where we define $F(\cdot)\equiv\partial_{X}f(X)\ast \cdot$, and $F(\cdot)$ is
linear to ``$\cdot$". So the above-mentioned definition of $\pi(X)$ fails to satisfy PI.
The problem arises from the derivative with respect to $X$. Therefore we redefine the derivative field as
\begin{equation}\label{def2}
\pi(X)\equiv M(\partial_{X})\varphi(X),\nonumber
\end{equation}
where $M(\cdot)$ is linear to ``$\cdot$" and has the covariant
transformation property
\begin{equation}
M(\cdot)\rightarrow M'(\cdot)=M(F(\cdot)).\nonumber
\end{equation}
Thus we have
\begin{eqnarray}
\pi'(X') &=&M'(\partial_{X'})\varphi'(X')\nonumber\\
 &=&M(F(\partial_{X'}))\varphi'(X')\nonumber\\
 &=&M(\partial_{X}f(X)\ast\partial_{X'})\varphi'(X')\nonumber\\
 &=&M(\partial_{X})\varphi(X)\nonumber\\
  &=&\pi(X).\nonumber
\end{eqnarray}
According to (\ref{eqn:PFIMBM}), $\pi(X)$ is now indeed a physical
field. We realize the principle of PI via the introduction of $M(\partial_X)$ and the replacement of
\begin{equation}\label{partial1}
\partial_{X}\rightarrow M(\partial _{X}).\nonumber
\end{equation}

\section{Background Matrix}\label{sec_BM}

The derivations in Section 2 are coordinate-free. If concrete calculations are concerned,
we may specify a coordinate system with basis $\{e_I\}$, and the completeness relation leads to
\begin{equation}
M(e^J)=e_Ie^I*M(e^J)=e_IM^{IJ}, \nonumber
\end{equation}
where $M^{IJ}\equiv e^I*M(e^J)$. We call $M^{IJ}$ the Background Matrix (BM) field.

If we choose another coordinate
frame $\{e_{I'}\}$, the coordinate transformation is $e_{I'}=T(e_I)=e_J e^J*T(e_I)=e_J T_{I}^{J}$,
with $T^J_I\equiv e^J*T(e_I)$ being the coordinate transformation matrix. Therefore, $M^{IJ}$ is accordingly transformed to
\begin{eqnarray}
M^{IJ}\rightarrow M^{I'J'}&=&e^{I'}*M(e^{J'})=T(e^I)*M(T(e^J))\nonumber\\
&=&(e^KT^I_K)*M(e^LT^J_L)\nonumber\\
&=&T^I_K(e^K*M(e^L))T^J_L\nonumber\\
&=&T^I_K M^{KL}T^J_L.\nonumber
\end{eqnarray}
So the component $M^{IJ}$ is coordinate-dependent and transforms in the same way as a common tensor.

Furthermore for $\partial_X$, we have
\begin{equation}
M(\partial _{X})=M(e^{J}\partial_{J})=M(e^{J})\partial_{J}=e_{K}M^{KJ}\partial_{J},\nonumber
\end{equation}
where the second and third items result from the facts that $\partial_{J}$ is a scalar operator and $M(\cdot)$ is linear to
``$\cdot$". So the replacement for the component $\partial_{J}$ of a multivector $\partial_{X}$ is
\begin{equation}\label{partial2}
\partial^{K}\rightarrow M^{KJ}\partial_{J}.\nonumber
\end{equation}

We now turn our attention to the physical implications of the BM and show that it contains LV and information
of anisotropy of spacetime. But ahead of that, for the integrity and
consistence of a complete framework, we further provide
another principle aside from the one of PI for the discussion
of covariant derivatives and various gauge fields related with local symmetries.

If a symmetry group is local to manifold, we must define a covariant derivative operator to maintain
the covariance of the Lagrangian under
gauge transformations. If there exists a scalar operator $D_J$, which applies to two arbitrary fields $\varphi_1(X)$ and $\varphi_2(X)$,
\begin{eqnarray} \label{Dproperty}
D_J(\varphi_1(X)\varphi_2(X))&=&D_J\varphi_1(X)\varphi_2(X)+\varphi_1(X)D_J\varphi_2(X), \nonumber\\
D_J\varphi(X) &=& \partial_J\varphi(X),\quad \textrm{ if }\varphi(X)\textrm{ is a scalar},
\end{eqnarray}
we demand the principle of covariance: Under the transformation
\begin{displaymath}
\varphi(X)\rightarrow R(X)\varphi(X), \quad \textrm{or} \quad \varphi(X)\rightarrow R(X)\varphi(X)R^{-1}(X),
\end{displaymath}
there is a corresponding transformation, $D_J\rightarrow D_J{'}$, such that
\begin{eqnarray}
D_J{'}(R(X)\varphi(X))&=&R(X)D_J\varphi(X), \nonumber\\
\textrm{or} \quad D_J{'}(R(X)\varphi(X)R^{-1}(X))&=&R(X)D_J\varphi(X)R^{-1}(X)\nonumber,
\end{eqnarray}
where $R(X)$ is an invertible multivector in $ \mathcal{G}$, standing for
various local symmetries, with specific matrix representations like SU(N) or SO(N). The operator $D_J$ is
named as covariant derivative, and the principle of covariance can
determine the forms of $D_J$ and further introduce gauge
fields with respect to the local symmetry $R(X)$ in the framework of geometric algebra.

$D_J$ in geometric algebra may have different forms in the Standard Model (SM), differential geometry,
and general relativity, just as a symmetry group may have different
representations. What we want to emphasize is that all definitions are equivalent and can be unified
in geometric algebra. For example, from the viewpoint of general
relativity, for a general multivector $\varphi(X)$, we have
\begin{eqnarray}
D_J \varphi(X)&=&D_J(\varphi^{K}e_{K})=D_J \varphi^{K}e_{K}+\varphi^{K}D_Je_{K}\nonumber\\
&=&\partial_J \varphi^{K}e_{K}+\varphi^{K}\Gamma_{JK}^Ie_{I}\nonumber\\
&=&(\partial_J \varphi^{I}+\Gamma_{JK}^I \varphi^{K})e_{I},\nonumber
\end{eqnarray}
where $D_Je_{K}=\Gamma_{JK}^Ie_{I}$,
with the coefficient $\Gamma_{JK}^I$ named as connection. Usually we define
$D_J\varphi^{I}\equiv \partial_J \varphi^{I}+\Gamma_{JK}^I\varphi^{K}$
as the covariant derivative in general
relativity. However in geometric algebra, we abandon this specific definition, and the property (\ref{Dproperty})
and the principle of covariance give an alternative choice.

Till now, we have all the physical and mathematical preparations ready.
Let us pause here and briefly sum up the basic ideas in our paper.
(i) When requiring the property of PI for an arbitrary field, we must introduce a
local matrix $M^{IJ}$ to modify a derivative $\partial^{I}$ to a co-derivative
$M^{IJ}\partial_{J}$. (ii) When promoting a global symmetry to a local one,
we have to change a derivative $\partial^{I}$ to a covariant derivative $D^{I}$ and acquire gauge fields.
Altogether these two considerations straightforwardly lead us to a new covariant co-derivative
operator $\partial^{I}\rightarrow M^{IJ}\partial_{J}\rightarrow M^{IJ}D_{J}$
or $\partial^{I}\rightarrow D^{I}\rightarrow M^{IJ}D_{J}$. This generation is the
essence for the origin of the LV terms in the SM.

When considering both the principles of PI and of covariance, we get
\begin{eqnarray}
M(D_{X})&=&M(e^{J}D_{J})=M(e^{J})D_{J}=e_{K}M^{KJ}D_{J},\nonumber
\end{eqnarray}
with the coordinate-free covariant multivector derivative
$D_{X}\equiv e^JD_J$, and $D_J$ being the component of $D_{X}$. So we arrive at
our replacement for the covariant multivector derivative $D_{X}$,
\begin{equation}\label{D1}
D_{X}\rightarrow M(D_{X}),
\end{equation}
and the replacement for its component $D_{J}$,
\begin{equation}\label{D2}
D^{K}\rightarrow M^{KJ}D_{J}.
\end{equation}

The replacements (\ref{D1}) and (\ref{D2}) are the consequences of the principles of both PI and
covariance. The first principle indicates the existence of the BM, and
the second is important to introduce covariant derivatives, local symmetries and gauge
fields. For the goal of this paper, to explore the BM and its physical implications, the principle
of PI is enough. But for completeness and clearness, we simply
discuss the principle of covariance. Now, we move on to spacetime, which can be part of general
geometric algebra space. So $X$ is replaced by spacetime coordinate $x$, and the indices are explicitly denoted by
$\alpha,\beta$ instead of $I,J$.

\section{Standard Model Supplement}\label{sec_SMS}

Section \ref{sec_BM} provides the essentials to construct a
mathematical-background-manifold-free and coordinate-free
framework for physics. One of the significant results is that in order to satisfy the principle of PI, the common derivative
$\partial_{\alpha}$ and covariant derivative $D_{\alpha}$ must be
generalized to $M^{\alpha\beta}\partial_{\beta}$ and
$M^{\alpha\beta}D_{\beta}$, with $M^{\alpha\beta}$ being the BM.
Except that, other basic fields remain untouched, because they do not involve with
these two derivatives. In this section, we follow this scheme and
focus on the physical implications and consequences from these new
introduced co-derivatives $M^{\alpha\beta}\partial_{\beta}$ and $M^{\alpha\beta}D_{\beta}$.

The effective Lagrangian of the minimal SM $\mathcal{L}_{\mathrm{SM}}$ consists of the following four parts
\begin{eqnarray}
\mathcal{L}_{\mathrm{SM}}&=& \mathcal{L}_{\mathrm{G}}+\mathcal{L}_{\mathrm{F}}+\mathcal{L}_{\mathrm{H}}+\mathcal{L}_{\mathrm{HF}},\nonumber\\
\mathcal{L}_{\mathrm{G}} &=& -\frac{1}{4}F^{a\alpha\beta } F_{\alpha\beta}^{a},  \label{eqn:SMG}\\
\mathcal{L}_{\mathrm{F}} &=&i\bar{\psi}\gamma^{\alpha}D_{\alpha}\psi,  \label{eqn:SMF} \\
\mathcal{L}_{\mathrm{H}}&=& (D^{\alpha}\phi)^{\dag}D_{\alpha}\phi + V(\phi). \label{eqn:SMHG}
\end{eqnarray}
Here $\psi$ is fermion field, $\phi$ is Higgs field, and $V(\phi)$ is its self-interaction.
$F_{\alpha\beta}^{a}\equiv\partial_{\alpha}A_{\beta}^{a}-\partial_{\beta}A_{\alpha}^{a}-gf^{abc}A_{\alpha}^{b}A_{\beta}^{c}$,
$D_{\alpha}\equiv\partial_{\alpha}+igA_{\alpha}$, and
$A_{\alpha}\equiv A_{\alpha}^{a}t^{a}$, with $g$ being the coupling constant,
$f^{abc}$ the structure constant, and $t^{a}$ the generator of gange groups respectively.
$\mathcal{L}_{\mathrm{HF}}$ is the Yukawa coupling between fermion and Higgs fields, which
is not related to $\partial^{\alpha}$ and $D^\alpha$, so it keeps
unchanged under the replacement (\ref{eqn:substitution}).
The chiral difference and the summation of chirality
and gauge index are omitted here for simplicity.

We divide $M^{\alpha\beta}$ into two parts
$M^{\alpha\beta}=g^{\alpha\beta}+\Delta^{\alpha\beta}$, with $g^{\alpha\beta}$ as the metric of spacetime.
(This decomposition will be fully discussed in the next section.)
Under (\ref{eqn:substitution}), the Lagrangians in (\ref{eqn:SMG}), (\ref{eqn:SMF}), and (\ref{eqn:SMHG}) become
\begin{eqnarray}
\mathcal{L}_{\mathrm{G}} &=& -\frac{1}{4}
 (M^{\alpha\mu}\partial_{\mu}A^{a\beta}-M^{\beta\mu}\partial_{\mu}A^{a\alpha}-gf^{abc}A^{b\alpha}A^{c\beta})   \nonumber\\
 &&\times(M_{\alpha\mu}\partial^{\mu}A_{\beta}^{a}-M_{\beta\mu}\partial^{\mu}A_{\alpha}^{a}-gf^{abc}A_{\alpha}^{b}A_{\beta}^{c}) \nonumber\\
  &=&-\frac{1}{4}F^{a\alpha\beta }F_{\alpha\beta}^{a}+ \mathcal{L}_{\mathrm{GV}},\label{eqn:SMSG}\\
\mathcal{L}_{\mathrm{F}} &=&i\bar{\psi}\gamma_{\alpha}M^{\alpha\beta}D_{\beta}\psi
  =i\bar{\psi}\gamma^{\alpha}D_{\alpha}\psi+  \mathcal{L}_{\mathrm{FV}}, \label{eqn:SMSF} \\
\mathcal{L}_{\mathrm{H}}&=&(M^{\alpha\mu}D_{\mu}\phi)^{\dag}M_{\alpha\nu}D^{\nu}\phi+ V(\phi)        \nonumber \\
  &=&(D^{\alpha}\phi)^{\dag}D_{\alpha}\phi + V(\phi)+ \mathcal{L}_{\mathrm{HV}},\label{eqn:SMSHG}
\end{eqnarray}
with the condition $M^{\alpha\beta}$ being real matrix to maintain the Lagrangian hermitian. The last three terms $\mathcal{L}_{\mathrm{GV}}$,
$\mathcal{L}_{\mathrm{FV}}$, and $\mathcal{L}_{\mathrm{HV}}$ in (\ref{eqn:SMSG}), (\ref{eqn:SMSF}), and (\ref{eqn:SMSHG})
are the supplement terms for the minimal SM, reading
\begin{eqnarray}
\mathcal{L}_{\mathrm{GV}}&=&
-\frac{1}{2}\Delta^{\alpha\beta}\Delta^{\mu\nu}(g_{\alpha\mu}\partial_{\beta} A^{a\rho}\partial_{\nu}A_{\rho}^{a}-\partial_{\beta}A_{\mu}^{a}
\partial_{\nu}A_{\alpha}^{a})  -F_{\mu\nu}^{a}\Delta^{\mu\alpha}\partial_{\alpha}A^{a\nu},\label{eqn:GV}  \\
\mathcal{L}_{\mathrm{FV}}&=&
 i\Delta^{\alpha\beta}\bar{\psi}\gamma_{\alpha}\partial_{\beta}\psi
  -g\Delta^{\alpha\beta}\bar{\psi}\gamma_{\alpha}A_{\beta}\psi, \label{eqn:FV} \\
\mathcal{L}_{\mathrm{HV}}&=& (g_{\alpha\mu}\Delta^{\alpha\beta}\Delta^{\mu\nu}+\Delta^{\beta\nu}+\Delta^{\nu\beta})
  (D_{\beta}\phi)^{\dag}D_{\nu}\phi.  \label{eqn:HGV}
\end{eqnarray}

Thus $\mathcal{L}_{\mathrm{SM}}$ is modified to an effective Lagrangian of the SM with
supplement terms (SMS) $\mathcal{L}_{\mathrm{SMS}}$,
\begin{eqnarray}
\mathcal{L}_{\mathrm{SMS}}= \mathcal{L}_{\mathrm{SM}} + \mathcal{L}_{\mathrm{LV}}, \nonumber
\end{eqnarray}
with
\begin{eqnarray}
\mathcal{L}_{\mathrm{LV}}\equiv\mathcal{L}_{\mathrm{GV}}+\mathcal{L}_{\mathrm{FV}}+\mathcal{L}_{\mathrm{HV}}.\nonumber
\end{eqnarray}
$\mathcal{L}_{\mathrm{SMS}}$ satisfies the
invariance of gauge group SU(3)$\bigotimes$SU(2)$\bigotimes$U(1)
and the invariance of PI. All the terms above in the
Lagrangians are Lorentz scalars at more fundamental level than
the minimal SM. But for the SM, $\Delta^{\alpha\beta}$ is treated as coupling
constants or background influences from this more fundamental
theory, and all the other fields for the SM are what we are studying, so
$\mathcal{L}_{\mathrm{LV}}$ is not Lorentz invariant under the observer's
Lorentz transformation on these fields. From this point of view, we call
the supplement term $\mathcal{L}_{\mathrm{LV}}$ the Lorentz invariance
violation term, and it contains the information of LV or anisotropy of spacetime in the SM.

To achieve a deeper insight and clearer understanding for the SMS here, let
us make a comparison with the commonly used Standard Model Extension
(SME)~\cite{SME98} and try to figure out the relations of the
various coupling constants. Keeping the conventions
in Ref. \refcite{SME98} and omitting detailed derivations, we
summarize our results in Table \ref{SMSandSME}.

\begin{table*}[h]
\tbl{Comparison of the Standard Model Supplement (SMS) and the Standard Model Extension (SME) in Ref. 6.
The notation $\langle\cdot\rangle$ means the vacuum expectation value. The subscripts $A$ and $B$ denote
the flavors of particles, and $G$, $W$, and $B$ mean SU(3), SU(2), and U(1) gauge fields respectively.}
{\begin{tabular}{cc}
\toprule
SMS&SME\\
\colrule
$\mathcal{L}_{\mathrm{FV}}$ & $\mathcal{L}_{\mathrm{lepton}}^{\mathrm{CPT}-\mathrm{even}}+\mathcal{L}_{\mathrm{lepton}}^{\mathrm{CPT}-\mathrm{odd}}$,
$\mathcal{L}_{\mathrm{quark}}^{\mathrm{CPT}-\mathrm{even}}+\mathcal{L}_{\mathrm{quark}}^{\mathrm{CPT}-\mathrm{odd}}$\\

$\mathcal{L}_{\mathrm{GV}}$ & $\mathcal{L}_{\mathrm{gauge}}^{\mathrm{CPT}-\mathrm{even}}+\mathcal{L}_{\mathrm{gauge}}^{\mathrm{CPT}-\mathrm{odd}}$\\

$\mathcal{L}_{\mathrm{HV}}$ & $\mathcal{L}_{\mathrm{Higgs}}^{\mathrm{CPT}-\mathrm{even}}+\mathcal{L}_{\mathrm{Higgs}}^{\mathrm{CPT}-\mathrm{odd}}$\\

$\langle\Delta_{\mu\nu}\rangle\delta_{AB}$ & $(c_L)_{\mu\nu AB}$, $(c_R)_{\mu\nu AB}$,
$(c_Q)_{\mu\nu AB}$, $(c_U)_{\mu\nu AB}$, $(c_D)_{\mu\nu AB}$\\

$g\langle\Delta_{\mu\nu}A^{\nu}\rangle\delta_{AB}$
& $(a_L)_{\mu AB}$, $(a_R)_{\mu AB}$, $(a_Q)_{\mu AB}$, $(a_U)_{\mu AB}$, $(a_D)_{\mu AB}$\\

$2\langle(g^{\gamma\rho}\Delta_{\gamma\beta}\Delta_{\rho\nu} g_{\alpha\mu}-\Delta_{\alpha\beta}\Delta_{\mu\nu})\rangle$
& $(k_{G})_{\beta\mu\nu\alpha}$, $(k_{W})_{\beta\mu\nu\alpha}$, $(k_{B})_{\beta\mu\nu\alpha}$\\

$4g^{\lambda\nu}\langle\partial_{\alpha}\Delta^{\mu\alpha}\rangle$
& $2(k_3)_{\kappa}\epsilon^{\kappa\lambda\mu\nu}$, $2(k_2)_{\kappa}\epsilon^{\kappa\lambda\mu\nu}$,
$4(k_1)_{\kappa}\epsilon^{\kappa\lambda\mu\nu}$, $2(k_{AF})_{\kappa}\epsilon^{\kappa\lambda\mu\nu}$\\
\botrule
\end{tabular}\label{SMSandSME}}
\end{table*}

We find: (i)~$\Delta^{\alpha\beta}$ provides the most equivalent coupling
constants in the SME of the LV items in the sectors of fermion, gauge,
and Higgs fields; (ii)~The various combinations of $\Delta^{\alpha\beta}$ as coupling constants own a
different CPT property. For example, $\Delta_{\mu\nu}$,
$\Delta_{\mu\nu}A^{\nu}$, $g^{\gamma\rho}\Delta_{\gamma\beta}\Delta_{\rho\nu}
g_{\alpha\mu}-\Delta_{\alpha\beta}\Delta_{\mu\nu}$, and
$\partial_{\alpha}\Delta^{\mu\alpha}$ are CPT-even, CPT-odd, CPT-even, and
CPT-odd respectively. The SME in Ref. \refcite{SME98} includes all the
possible LV terms of spontaneous symmetry breaking
for the SM and it is mentioned that all these LV terms may origin from a
fundamental theory. Thus what we perform in this paper shows a fundamental way for the
LV terms in the SM from basic principles.

\section{Lorentz Invariance Violation Matrix}\label{sec_LVM}

Now let us turn to the local BM $M^{\alpha\beta}$, of which the vacuum expectation value is used for the coupling constants in
(\ref{eqn:GV}), (\ref{eqn:FV}) and  (\ref{eqn:HGV}). We decompose $M^{\alpha\beta}$ into two parts
\begin{equation}\label{eqn:Mdivide}
M^{\alpha\beta}=g^{\alpha\beta}+\Delta^{\alpha\beta},\nonumber
\end{equation}
where $g^{\alpha\beta}$ is the metric of spacetime. Since all the elements of $M^{\alpha\beta}$ or $\Delta^{\alpha\beta}$
are dimensionless, they naturally encode the strength of LV or the degree of  anisotropy
of spacetime~\cite{Liberati2009}, i.e.
\begin{equation}\label{LIMmeaning}
\Delta^{\alpha\beta}\quad
\left\{
  \begin{array}{ll}
           = 0,           & \quad \textrm{no LV}, \\
    \rightarrow 0,        & \quad \textrm{small LV}, \\
    = \textrm{otherwise}, & \quad \textrm{large LV}. \\
  \end{array}
\right.\nonumber
\end{equation}
Hence we call $\Delta^{\alpha\beta}$ the Lorentz invariance
Violation Matrix (LVM), and its entries will be constrained with the help of laboratory experiments~\cite{ge} and astronomical observations~\cite{am98,el00,m05,as09,emn09,XiaoLIV2,Shao}.
Generally speaking, $\Delta^{\alpha\beta}$ depend on the types of particles. While $\varphi(x)$ can be re-scaled to absorb one
of the 16 degrees of freedom in $\Delta^{\alpha\beta}$, so that only 15 are left physical.
Thus in this paper, we assume $M^{00}=g^{00}$, or $\Delta^{00}=0$.

As a result of the LVM, we may attain various modified dynamical equations of fields, as well as dispersion relations
from the effective Lagrangian $\mathcal{L}_{\mathrm{SMS}}$. Here as a preliminary test of our
construction, we take the Dirac equation for free fermion field $\psi(x)$ as an example.
First, we replace $\partial^\alpha$ to $M^{\alpha\beta}\partial_\beta$,
\begin{equation}\label{eqn:Dirac}
(i\gamma_{\alpha}M^{\alpha\beta}\partial_{\beta}-m)\psi(x)=0.\nonumber
\end{equation}
Second, we multiply $(i\gamma_{\alpha}M^{\alpha\beta}\partial_{\beta}+m)$ on both sides,
\begin{equation}
(g_{\alpha\mu}M^{\alpha\beta}M^{\mu\nu}\partial_{\beta}\partial_{\nu}+m^2)\psi(x)=0.\nonumber
\end{equation}
With the Fourier transformation $\psi(x)=\int\psi(p)e^{-ip\cdot x}dp$, the extended dispersion relation becomes
\begin{equation}\label{eqn:disper}
p^2+g_{\alpha\mu}\Delta^{\alpha\beta}\Delta^{\mu\nu}p_{\beta}p_{\nu}+2\Delta^{\alpha\beta}p_{\alpha}p_{\beta}=m^2.
\end{equation}
We see that the left-hand side of (\ref{eqn:disper}) is not invariant under the observer's Lorentz
transformation on $p$, reflecting the influences from the
fundamental theory. So we claim that the last two items of the left-hand side of (\ref{eqn:disper}), which are the
extensions of the ordinary mass-energy relation $p^2=m^2$, contain the information of LV.

Systematic LV effects from the general form of $\Delta^{\alpha\beta}$ still need further studies, but
here we merely employ a special SO(3) invariant model of LVM to demonstrate our mechanism. So we assume
\begin{equation}\label{eqn:LIMsimple}
\Delta^{\alpha\beta}=\left(
                       \begin{array}{cccc}
                         0 & 0  & 0  & 0  \\
                         0 &\xi & 0  & 0  \\
                         0 & 0  &\xi & 0  \\
                         0 & 0  & 0  &\xi \\
                       \end{array}
                     \right).
\end{equation}
Substituting (\ref{eqn:LIMsimple}) into (\ref{eqn:disper}) gives the extended dispersion relation
for free fermion field in this simple case,
\begin{eqnarray}\label{fermionDisper}
E^2=(1-\delta)\vec{p}^2+m^2, \quad \delta\equiv 2\xi-\xi^2.
\end{eqnarray}

\section{Comparison with Experimental Data}\label{sec_EX}

We could utilize proton to determine the upper bound of $\xi$.
The photopion production of nucleon in the Greisen-Zatsepin-Kuz'min (GZK) cutoff~\cite{G,ZK} observations gives an
available energy threshold $E\approx{10^{19}}$~eV (see, e.g., Ref. \refcite{Wolfgang}).
The dominant channel for this production begins with $p+\gamma\rightarrow\Delta^+~(1232\textrm{~MeV})$.
We concentrate on the head-on collision of the proton in cosmic rays and the
photon from the Cosmic Microwave Background (CMB). The dispersion relations for $p$, $\gamma$,
and $\Delta^+$ are similar as that in (\ref{fermionDisper}), with the corresponding $\delta$'s denoted by
$\delta_{p}$, $\delta_{\gamma}$, and $\delta_{\Delta^+}$. As considering the LV of the high energy protons from cosmic rays,
we are allowed to assume $\delta_{\gamma}=\delta_{\Delta^+}=0$. So in this channel, we have
\begin{equation}
p_{\Delta^+}^2 \leqslant (p_{p}+p_{\gamma})^2,\nonumber
\end{equation}
with $p_{p}=(E_{p},\vec{p}_{p})$, $p_{\gamma}=(\omega, \vec{p}_{\gamma})$, and $p_{\Delta^+}=(E_{\Delta^+}, \vec{p}_{\Delta^+})$ being the 4-momenta of $p$, $\gamma$, and $\Delta^+$ respectively. Using (\ref{fermionDisper}), we obtain
\begin{eqnarray}
m_{\Delta^+}^2 &\leqslant& (E_{p}+\omega)^2 - (\vec{p}_{p}+\vec{p}_{\gamma})^2 \nonumber\\
&=&E_{p}^2-\vec{p}_{p}^2+\omega^2-\vec{p}_{\gamma}^2+2\omega E_{p}-2\vec{p}_{p}\cdot\vec{p}_{\gamma} \nonumber\\
&=&m_{p}^2-\delta_{p}\vec{p}_{p}^2+2\omega E_{p}-2\vec{p}_{p}\cdot\vec{p}_{\gamma}\nonumber\\
&=&m_{p}^2-2\xi_{p} E_{p}^2 + 4\omega E_{p}.\nonumber
\end{eqnarray}
For high energy protons, $E_{p}^2\simeq \vec{p}_{p}^2$, and
$\vec{p}_{p}\cdot\vec{p}_{\gamma}=-\omega E_{p}$ due to the
head-on collision. We keep only $2\xi_p$ in $\delta_p$ since the quadratic term $-\xi_p^2$ is negligible, so the final expression for the energy $E_{p}$ of high energy protons satisfies the inequality
\begin{equation}\label{protonE}
2\xi_{p} E_{p}^2 - 4\omega E_{p} + m_{\Delta^+}^2-m_{p}^2 \leqslant 0.\nonumber
\end{equation}
In case of no LV for high energy protons, namely $\xi_{p}=0$, we
have
$E_{p}\geqslant(m_{\Delta^+}^2-m_{p}^2)/(4\omega)=5.3\times10^{19}$~eV,
which is the common threshold energy for the GZK cutoff. (Here
$m_{\Delta^+}=1232$~MeV, $m_{p}=938$~MeV, and the mean energy of the
photons in the CMB is taken as $\bar{\omega}\simeq
6\times10^{-4}$~eV.) For high energy photons from the CMB, we take
$\omega=5\bar{\omega}=3$~meV for calculations. A small positive
$\xi_{p}$ will increase the threshold energy, which can be higher
than $5.3\times10^{19}$~eV, and the constraint on $\xi_{p}$ is
\begin{equation}
\xi_{p} \leqslant \frac{2\omega^2}{m_{\Delta^+}^2-m_{p}^2}=2.8 \times 10^{-23}.\nonumber
\end{equation}
This constraint is consistent with our previous estimate in Ref. \refcite{XiaoLIV1}.

\section{Conclusion}

With a general requirement of the  physical independence or physical
invariance of mathematical background manifold, we introduce the
background matrix $M^{\alpha\beta}$, and the replacement of the common derivative operators by the
covariant co-derivative ones. This replacement gives rise to supplement terms in
the minimal standard model. We introduce a Lorentz invariance violation matrix
$\Delta^{\alpha\beta}$, which is able to characterize the Lorentz invariance
violation or spacetime anisotropy. Thus we have a feeling that the principles of physical invariance and
covariance are more fundamental than Lorentz invariance or spacetime isotropy.

\section*{Acknowledgements}

This work is partially supported by National Natural Science Foundation of China (No. 10721063 and No. 10975003), by the Key
Grant Project of Chinese Ministry of Education (No. 305001),
and by the Research Fund for the Doctoral Program of Higher Education (China).

\end{document}